\documentclass[11pt]{article}

\usepackage{acl}

\usepackage{subcaption}
\usepackage{times}
\usepackage{latexsym}
\usepackage{booktabs}
\usepackage{multirow}
\usepackage{array}
\usepackage[linesnumbered,ruled,vlined]{algorithm2e}
\usepackage{wrapfig}
\usepackage{colortbl,xcolor}
\usepackage{pifont}
\usepackage{amssymb}
\newcommand{\cmark}{\textcolor{green!70!black}{\ding{51}}} 
\newcommand{\xmark}{\textcolor{red}{\ding{55}}}            
\usepackage[most]{tcolorbox}
\tcbset{
    mypromptbox/.style={
        enhanced,
        colback=white,
        colframe=gray!40,
        coltitle=white,
        fonttitle=\bfseries\sffamily,
        arc=3mm,
        boxrule=0.5pt,
        left=5pt, right=5pt, top=3pt, bottom=3pt,
        boxsep=3pt,
        borderline={0.5pt}{0pt}{gray!20},      
        borderline={1pt}{2pt}{blue!50!black}, 
        attach boxed title to top left,
        boxed title style={colback=blue!50!black, colframe=blue!50!black, arc=1.5mm, boxrule=0pt},
        fontupper=\small,                      
        fonttitle=\small\bfseries\sffamily     
    }
}

\usepackage{microtype}
\usepackage{enumitem}

\newtcolorbox{promptbox}[1]{%
  enhanced,
  colback=white,
  colframe=black!55,
  boxrule=0.8pt,
  arc=8pt,               
  left=3mm,right=3mm,top=10mm,bottom=2mm,
  overlay={%
    \node[
      anchor=west,
      fill=black!75,
      text=white,
      font=\bfseries\sffamily,
      rounded corners=8pt,
      inner xsep=10pt,
      inner ysep=6pt
    ] at ([xshift=6mm,yshift=-6mm]frame.north west) {#1};
  },
}

\newtcolorbox{casebox}[2]{%
  enhanced,
  sharp corners,
  boxrule=0.6pt,
  colback=#1!4,
  colframe=#1!70!black,
  coltitle=white,
  fonttitle=\bfseries,
  title={#2},
  attach boxed title to top left={xshift=2mm,yshift=-2mm},
  boxed title style={
    colback=#1!70!black,
    sharp corners,
    boxrule=0pt,
    left=6pt,right=6pt,top=2pt,bottom=2pt
  },
  left=6pt,right=6pt,top=6pt,bottom=6pt,
  before skip=8pt, after skip=8pt,
}
\newcommand{\field}[1]{\textbf{#1}\hspace{0.4em}}

\usepackage[T1]{fontenc}

\usepackage[utf8]{inputenc}

\usepackage{inconsolata}

\usepackage{graphicx}

\title{M$^2$Note: Continual Evolution of Vision Language Models via Mistake Notebook Learning}

\author{Haiwen Li$^1$\thanks{Work done during the internship at AMAP, Alibaba.} \quad
Jing Tang$^2$\thanks{Project Lead.} \quad
Rui Chen$^2$ \quad
Lei Sun$^2$ \quad
Zining Chen$^2$ \quad
Xiangxiang Chu$^2$ \\
$^1$Beijing University of Posts and Telecommunications\\
$^2$AMAP, Alibaba Group
}

\begin{document}
\maketitle
\begin{abstract}
Vision Language Models (VLMs) have demonstrated remarkable capabilities in multimodal reasoning tasks, yet they still suffer from recurring failures, such as skipping key visual checks, misapplying domain rules, and hallucinating unsupported concepts. Most existing solutions rely on supervised fine-tuning (SFT) and reinforcement learning (RL), which are expensive to iterate and can be brittle under distribution shift. To this end, we propose \textbf{M}ultimodal \textbf{M}istake \textbf{Note}book Learning (M$^2$Note), a training-free continual evolution framework that externalizes learning into an editable memory. M$^2$Note transforms failed trajectories into compact subject–guidance notes: the subject summarizes the underlying domain and concept, while the guidance provides actionable verification steps that can be reused in future inference. At test time, M$^2$Note retrieves relevant notes via multimodal retrieval-augmented generation (RAG) and appends them to the model context, steering reasoning away from previously observed pitfalls. To stabilize continual evolution, we adopt batch-level post-verification with rollback, which commits notebook edits only if they improve performance on the same batch, reducing noisy updates and preventing regressions. M$^2$Note supports both \textit{self-evolving}, where the same VLM acts as solver and supervisor, and \textit{cross-model evolving}, where a stronger supervisor guides a weaker solver, enabling capability transfer without weight updates. Experiments on six multimodal reasoning benchmarks show consistent improvements across domains and backbones, while achieving strong cost and sample efficiency and remaining complementary to Chain-of-Thought (CoT) prompting. The code is available at \href{https://github.com/AMAP-ML/M2Note}{\textit{\color{blue!37}https://github.com/AMAP-ML/M2Note}}.
\end{abstract}

\begin{figure*}
\centering
\includegraphics[width=0.8\textwidth]{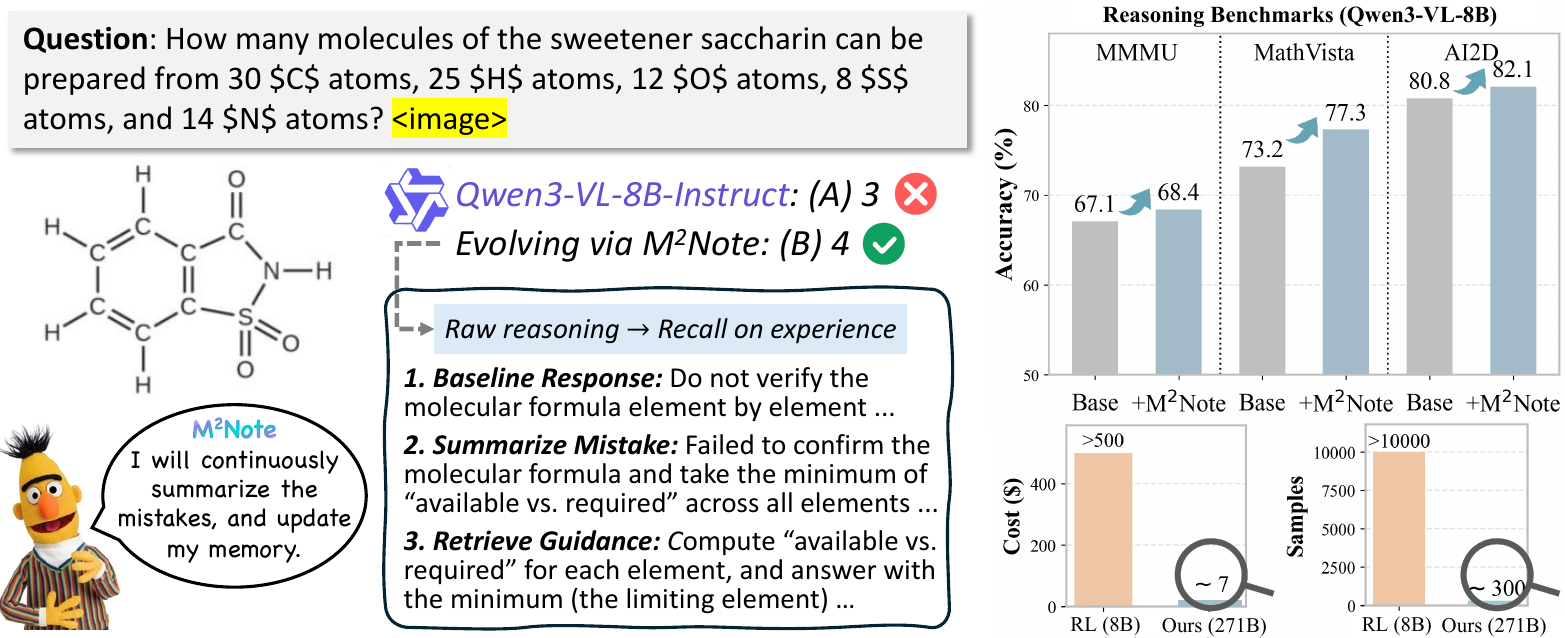}    
\caption{\textbf{Overview of M$^2$Note}. (Left) The system updates an external mistake notebook from incorrect responses and retrieves task-relevant guidance at inference time to refine reasoning, for instance, limiting-reagent counting for saccharin. (Right) Accuracy gains on different benchmarks (e.g., MMMU~\cite{mmmu}, MathVista~\cite{mathvista}, and AI2D~\cite{ai2d}) via M²Note, together with cost- and sample-efficiency comparisons.}
\label{imgs:intro}
\end{figure*}

\section{Introduction}
Vision Language Models (VLMs)~\cite{vlm, hurst2024gpt, qwen3-vl, seedvl, llava} have become a general interface for multimodal tasks, including STEM reasoning~\cite{mmmu,mathvista,mathvision,wemath}, chart and diagram understanding~\cite{chartqa,ai2d}, document OCR~\cite{ocrbench,ccocr}, visual question answering~\cite{mmbench,mmstar,realworldqa,univg-r1}, and video understanding~\cite{mvbench,mmvu,videostar}. Despite strong progress, VLMs in realistic settings still exhibit repeatable failure modes, such as missing key visual evidence or over-relying on superficial cues. Correcting such behaviors efficiently, without sacrificing robustness or requiring costly retraining, remains a practical challenge.

The dominant adaptation paradigm is parameter-based post-training, such as supervised fine-tuning (SFT)~\cite{llava,sft} and reinforcement learning (RL)~\cite{ppo,dpo,grpo,dapo}. Although effective, these methods are expensive, slow to iterate, and prone to regression. More importantly, once model weights are updated, test-time behavior becomes fixed, making continual improvement difficult in dynamic environments. This motivates growing interest in training-free adaptation through in-context steering.

Existing training-free approaches mainly fall into two categories: \ding{182} prompt optimization methods~\cite{llmrhuman,llmrop,autoprompt} refine global instructions, but often provide advice that is too coarse to target recurring errors across diverse multimodal tasks; \ding{183} memory-based methods~\cite{expel,incontextprinciple,reflexion,zhang2025agentic} store instance-level experiences for retrieval, but they often lack abstraction, leading to redundant memories and limited generalization beyond superficially similar cases. These limitations are especially pronounced in multimodal settings, where visual inputs are highly diverse and sparse.

To address this, we propose \textbf{M}ultimodal \textbf{M}istake \textbf{Note}book Learning (M$^2$Note), a tuning-free framework, which extends the learning-from-mistakes paradigm~\cite{incontextprinciple,mnl} to VLMs. M$^2$Note maintains an external mistake notebook that stores structured \textit{subject-guidance} pairs and retrieves relevant notes via multimodal embeddings~\cite{qwen3-vl-embedding,gme} to augment the model context. Given a multimodal query, the VLM first recalls relevant notes and produces an answer. If the answer is incorrect, a reflective supervisor summarizes the failure into a high-level subject and concise actionable guidance, then writes or merges the resulting note into the notebook. In this way, recurring errors are distilled into compact, reusable experience, shifting adaptation from model parameter updates to semantic updates of an external notebook.

Beyond being training-free, M$^2$Note is designed as a closed-loop evolution protocol:
\textit{generate $\rightarrow$ reflect $\rightarrow$ update $\rightarrow$ verify}. 
Whenever notebook updates are proposed, the system re-evaluates the same batch with the updated notebook and accepts the change only if performance improves; otherwise, it rolls back. This conservative \textit{accept-if-improves} mechanism stabilizes continual evolution and suppresses noisy updates. The framework supports both \textit{self-evolving}, where the same VLM acts as solver and supervisor, and \textit{cross-model evolving}, where a stronger VLM supervises a weaker one to transfer knowledge without any parameter updates.

Experiments on six multimodal reasoning benchmarks show that M$^2$Note consistently improves VLM performance. Compared with RL-based self-evolving methods such as VisPlay~\cite{visplay}, EvoLMM~\cite{evolmm}, and Vision-Zero~\cite{visionzero}, M$^2$Note achieves strong gains with substantially lower training cost and fewer samples, since it evolves only a compact external notebook rather than model parameters. We introduce the following technical components:
\begin{itemize}
\item We present M$^2$Note, a multimodal mistake notebook learning framework that improves VLM reasoning by storing and retrieving subject-level guidance in an external memory.
\item We propose a stable closed-loop evolution protocol with batch-level verification, supporting both self-evolving and cross-model evolving settings while remaining tuning-free.
\item We deliver consistent gains across six multimodal reasoning benchmarks, with strong cost and sample efficiency compared with SFT- and RL-based methods.
\end{itemize}

\section{Related Work}

\textbf{VLMs and Multimodal Reasoning.}
Multimodal large language models (MLLMs)~\cite{radford2021learning, qwen3-vl, liu2024improved, wang2025internvl3, hurst2024gpt, comanici2025gemini, finger, beyondquerybias} extend LLMs with visual perception and have achieved strong performance on multimodal understanding and reasoning tasks. Benchmarks such as MMMU~\cite{mmmu}, MathVista~\cite{mathvista}, AI2D~\cite{ai2d}, ChartQA~\cite{chartqa}, MMStar~\cite{mmstar}, and RealworldQA~\cite{realworldqa} evaluate challenging abilities including STEM reasoning, diagram understanding, and real-world visual reasoning.

\textbf{Self-Evolving and Training-Free Adaptation.}
Recent self-evolving methods improve models via closed-loop interaction, often by assigning co-evolving roles (e.g., a challenger and a solver) to the same base model. Representative examples~\cite{rzero,visionzero,visplay,rise,dpe} use self-play to generate increasingly informative training signals and typically internalize the resulting improvements through parameter updates, such as supervised fine-tuning~\cite{sft,chord} and reinforcement learning~\cite{ppo,grpo,dagrpo,treegrpo}. While effective, these methods are costly and can be brittle under distribution shift~\cite{chen2023fireact,zeng2024agenttuning,zhai2025agentevolver}. Training-free alternatives instead adapt models through inference-time context, primarily including prompt optimization~\cite{chain,autoprompt,llmrhuman,llmrop} and memory-augmented inference~\cite{reflexion, expel,memento,zhang2025agentic,training-free-grpo,reasoningbank}. Within the latter, learning-from-mistakes methods~\cite{self-refine,incontextprinciple, expel,mnl} convert failures into reusable vexperience for future inference. Our work extends this paradigm to VLMs by distilling multimodal failures into compact subject-level guidance and updating memory with batch-level verification.

\section{Methodology}
\subsection{Overview}
We propose \textbf{M}ultimodal  \textbf{M}istake  \textbf{Note}book Learning (M$^2$Note), a training-free framework that improves VLMs by maintaining an external mistake notebook. Unlike prior context-optimization methods mainly developed for text-only settings~\cite{mnl,synapse,agentworkflowmemory,incontextprinciple,memento}, M$^2$Note extends the \textit{learning-from-mistakes}~\cite{mnl,incontextprinciple} paradigm to multimodal tasks.

\begin{figure*}[t]
\centering
\includegraphics[width=\textwidth]{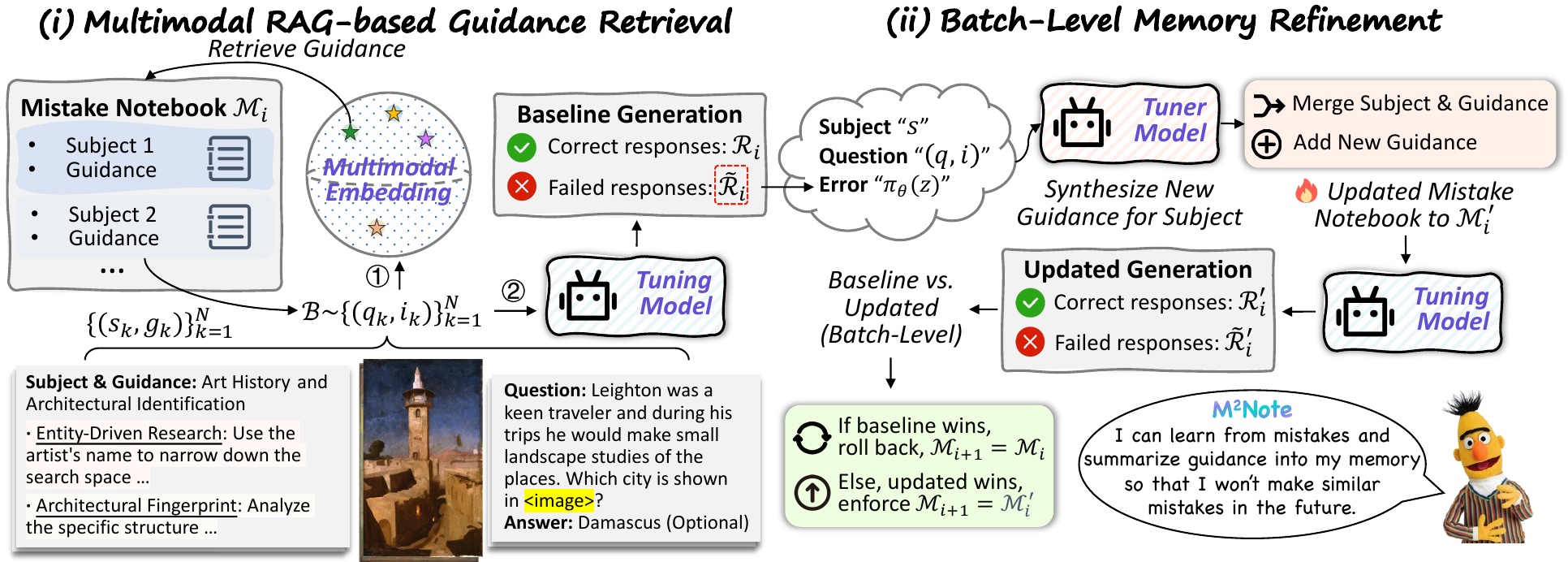}    
\caption{\textbf{The M$^2$Note evolving protocol.} M$^2$Note improves a VLM through a closed loop with two roles: a Tuning Model for solving multimodal queries with retrieved notebook guidance, and a Tuner Model for analyzing mistakes and refining the notebook. The process consists of (i) \textbf{Multimodal RAG-based Guidance Retrieval}, where the Tuning Model retrieves relevant subject-guidance notes from memory to generate responses, and (ii) \textbf{Batch-level Memory Refinement}, where the Tuner Model summarizes failures into new or merged notes, after which the update is verified and accepted only if it improves batch-level performance.}
\label{imgs:main}
\end{figure*}

As illustrated in Figure~\ref{imgs:main}, M$^2$Note involves two roles: a \textbf{Tuning Model} ($\pi_{\theta}$), which solves multimodal problems with retrieved notebook guidance, and a \textbf{Tuner Model} ($\pi_{tuner}$), which analyzes the Tuning Model’s errors and writes reusable guidance back into the notebook. The notebook is initialized as an external memory $\mathcal{M}$ that stores structured triplets $(s,g,e)$, where $s$ denotes the subject, summarizing the underlying task and domain, $g$ denotes the corresponding guidance, i.e., concise recommendations associated with that subject, and $e \in \mathbb{R}^d$ represents the embedding produced by a multimodal embedding model for retrieval. Examples of the notes are provided in Appendix~\ref{appendix:vis}.

The framework runs in a closed loop with three steps:  
\ding{182} the Tuning Model retrieves relevant notes from $\mathcal{M}$ and generates baseline responses;  
\ding{183} the Tuner Model summarizes failed cases into new subject-guidance notes, and decides whether to add or merge them;  
\ding{184} the system reruns the same batch with the updated memory and accepts the update only if the new memory improves batch-level performance; otherwise, it rolls back. 

\subsection{Problem Formulation}
M$^2$Note can be formulated as a context optimization problem: instead of updating the parameters of $\pi_{\theta}$, we refine only the external memory to maximize the expected reward, ensuring that the retrieved guidance provides effective assistance. Formally, let $\mathcal{D}$ denote a data distribution over multimodal queries $(q,i)$, optionally paired with $y$ (when the training split is available), where $q$ denotes the text question, $i$ the associated image, and $y$ the optional ground-truth answer when available. We seek an optimal memory $\mathcal{M}^*$:
\begin{equation}
\begin{aligned}
&\mathcal{M}^{*}
= \arg\max_{\mathcal{M}} \; \mathbb{E}_{(q,i)\sim\mathcal{D}}
\Big[ R\big(\pi_{\theta}(z)\big) \Big],\\
&\text{where}~z = (q,i) \oplus \mathrm{Ret}((q,i),\mathcal{M}),
\end{aligned}
\end{equation}
where $\oplus$ is context concatenation, and $\mathrm{Ret}(\cdot,\cdot)$ represents retrieving guidance from the notebook. We consider two instantiations of the reward function $R(\cdot)$:  
\ding{182} \textbf{Supervised setting.} If ground-truth labels $y$ are available, the model response is first parsed into a final prediction, and $R(\cdot)$ is instantiated as a binary reward by comparing the parsed prediction against the ground truth, i.e., it returns $1$ if they match and $0$ otherwise; \ding{183} \textbf{Label-free setting.} If labels are unavailable, $R(\cdot)$ adopts an LLM-as-a-Judge~\cite{llm-as-a-judge}, which evaluates whether the response is correct given the input (and task instruction if available), and outputs a binary judgment as reward. In both cases, the optimization process is delegated to $\pi_{tuner}$, which identifies failures, summarizes guidance, and updates the memory.

\subsection{The M$^2$Note Evolution Protocol}
As illustrated in Figure~\ref{imgs:main}, M$^2$Note follows a closed-loop protocol that iteratively retrieves guidance, updates memory from mistakes, and verifies whether the update should be kept.

\textbf{Multimodal RAG-based Guidance Retrieval}.  
The Tuning Model $\pi_{\theta}$ uses the current memory $\mathcal{M}_i$ in a multimodal RAG manner~\cite{rag,mrag}. For the $i$-th iteration, given an input batch $\mathcal{B}_i=\{(q_k,t_k)\}_{k=1}^N$, each query $(q_k,t_k)$ is fed into a multimodal embedding model $\mathcal{E}$ to obtain an embedding $e_k = \mathcal{E}(q_k,i_k) \in \mathbb{R}^d$. The system then computes the cosine similarity between $e_k$ and each note embedding in memory, ranks all notes by cosine similarity, and returns the top-$K$ results (the following shows $K=1$):
\begin{equation}
\begin{aligned}
(s_k,g_k)
&= \arg\max_{(s,g)} \left\{
\frac{\langle e_k, e\rangle}{\lVert e_k\rVert\,\lVert e\rVert}
\;\middle|\; (s,g,e)\in \mathcal{M}_i,\right. \\
&\qquad\qquad\left.
\frac{\langle e_k, e\rangle}{\lVert e_k\rVert\,\lVert e\rVert} > \alpha
\right\}
\end{aligned}
\end{equation}
where $\langle e_k, e\rangle$ denotes the inner product; $\lVert\cdot\rVert$ denotes the $\ell_2$ normalization; and $\alpha$ controls the relevance of the retrieved content. 

The retrieved notes are concatenated as advisory context to the VLM input, i.e., $z_k=(q_k, i_k) \oplus (s_k, g_k)$. In addition, the Tuning Model is instructed to incorporate the extra context critically rather than follow it blindly, which helps to reduce hallucinations (Appendix~\ref{appendix:hallu}, Table~\ref{tab:huanjue} and Figure~\ref{img:huanjue}). The initial response, denoted as $\pi_{\theta}(z_k)$, is fed into the reward function $R(\cdot)$. Aggregating and partitioning the responses within the batch yields two mutually exclusive sets: correct responses are collected into $\mathcal{R}_i=\{k~|~R(\pi_{\theta}(z_k))=1\}$, while failed trajectories form $\tilde{\mathcal{R}}_i=\{k~|~R(\pi_{\theta}(z_k))=0\}$. The batch accuracy is then computed as the performance baseline for the current iteration.

\textbf{Batch-Level Memory Refinement}.  
Given the failure set $\tilde{\mathcal{R}}_i$, the Tuner Model $\pi_{\text{tuner}}$ summarizes reusable guidance and proposes notebook updates. An update is accepted only if it improves batch-level performance; otherwise, it is rolled back.

(1) \textit{Subject Extraction $\sigma$.} For each failed sample index $k \in \tilde{\mathcal{R}}_i$, the system infers a \textit{subject} that summarizes the underlying concept of the query. Concretely, as shown below, we apply a prompted operator $\sigma(\cdot)$ implemented by $\pi_{\text{tuner}}$:
\begin{equation}
s'_k = \pi_{\text{tuner}}(\sigma(q_k, i_k)), \quad k \in \tilde{\mathcal{R}}_i.
\end{equation}
The subject is expected to be domain-aware (e.g.,
math, chemistry, or diagram understanding), specific enough to capture recurring error patterns, but not tied to instance-specific details (e.g., exact entities or numbers).

(2) \textit{Guidance Synthesis $\phi$.} Given a failed trajectory, $\pi_{\text{tuner}}$ further generates a guidance that is directly actionable at inference time. The guidance is distilled from the failed response $\pi_{\theta}(z_k)$ and the available feedback, abstracting the cause of failure into verification-oriented steps:
\begin{equation}
g'_k = \pi_{\text{tuner}}(\phi((q_k, i_k), \pi_\theta(z_k))),
~ k \in \tilde{\mathcal{R}}_i,
\end{equation}
where $\phi(\cdot)$ is a prompted “mistake-to-guidance” operator. In practice, $g'_k$ is constrained to be concise, structured, and oriented toward checks and invariants (e.g., “use the artist's name to narrow down the search space” in Figure~\ref{imgs:main}).

(3) \textit{Subject Merging and Memory Update $\mu$.}  
To avoid fragmentation, M$^2$Note optionally merges new notes with existing ones. For each candidate note $(s'_k,g'_k)$, the tuner predicts an edit action:
\begin{equation}
a_k = \pi_{\text{tuner}}(\mu(s'_k, s_k)) \in \{\texttt{add},\texttt{merge}\}.
\end{equation}
If $a_k=\texttt{Merge}$, $\pi_{\text{tuner}}$ synthesizes the merged subject $s_k^{*}$ and the merged guidance $g_k^{*}$; otherwise, they are set to $s'_k$ and $g'_k$, respectively. For each newly created or merged note, we compute its embedding as $e_k^*=\mathcal{E}(s_k^*\oplus g_k^*)$. The updated memory $\mathcal{M}'_{i}$ is obtained by incorporating all entries $\{(s_k^*, g_k^*, e_k^*)\}_{k=1}^N$ derived from this batch of data.

(4) \textit{Batch-Level Post-Verification.}
After proposing $\mathcal{M}_{i}'$, we rerun the same batch using the updated memory. The batch-level accuracy before and after the update is computed by the formula:
\begin{equation}
\text{Acc}(\mathcal{B}_i|\mathcal{M})=\frac{1}{N}\sum_{k=1}^{N} R(\pi_{\theta}(z_k|\mathcal{M})).
\end{equation}
We accept the update if it improves performance:
\begin{equation}
\mathcal{M}_{i+1}=
\begin{cases}
\mathcal{M}'_{i}, & \text{if } \text{Acc}(\mathcal{B}_i|\mathcal{M}'_i)\ge \text{Acc}(\mathcal{B}_i|\mathcal{M}_i),\\
\mathcal{M}_{i}, & \text{otherwise (rollback)}.
\end{cases}
\end{equation}
This “accept-if-improves” rule enforces monotonic non-degradation at the batch level, making notebook evolution robust to occasional low-quality reflections. It also acts as an implicit regularizer that curbs uncontrolled notebook growth, since only useful notes survive. The pseudo-code of the overall procedure is shown in Algorithm 1, and all prompt templates are provided in Appendix~\ref{appendix:prompt}.

\begin{algorithm}[t]
\SetCommentSty{textit}
\SetKwComment{tcp}{// }{}
\small
\caption{M$^2$Note evolving protocol}
\KwIn{Tuning Model $\pi_\theta$; Tuner Model $\pi_{tuner}$; Data Corpus $\mathcal{D}$; Reward function $R(\cdot)$; RAG top-$K$ and $\alpha$}
\KwOut{Mistake Notebook $\mathcal{M}$}
Initialize $\mathcal{M}_{0}\leftarrow\emptyset$ \;
\For{each $\mathcal{B}_i=\{(q_k,i_k)\}_{k=1}^N$ sampled from $\mathcal{D}$}{
  \tcp{(1) Retrieve guidance and run baseline}
  \For{each $k \in \mathcal{B}_{i}$}{
    $(s_k,g_k)\leftarrow \mathrm{Ret}((q_k,i_k),\mathcal{M}_i;K,\alpha)$ \;
    $z_k=(q_k,i_k)\oplus(s_k,g_k)$ \;
  }
  $\mathrm{Acc}_{base}\leftarrow \frac{1}{N}\sum_{k=1}^{N} R(\pi_{\theta}(z_k|\mathcal M_i))$ \;

  \tcp{(2) Propose notebook edits from failures}
  Backup memory $\mathcal{M}_i'\leftarrow \mathcal{M}_i$ \;
  \For{each $k$ such that $R(\pi_{\theta}(z_k|\mathcal M_i)=0$}{
    $\pi_{tuner}$ generates a subject-guidance pair $(s_k',g_k')$ \ and compares $s_k$ and $s_k'$ to make an action $a_k$ \;
    \uIf{$a_k=\texttt{add}$}{
      $\mathcal{M}_i'\leftarrow \mathcal{M}_i'\cup(s_k',g_k',\mathcal{E}(s_k'\oplus g_k'))$
    }
    \Else{
      $\pi_{tuner}$ merges $s_k \oplus s_k'$ into $s_k^*$ and $g_k \oplus g_k'$ into $g_k^*$ \;
      $\mathcal{M}_i'\leftarrow \mathcal{M}_i'\cup(s_k^{*},g_k^{*},\mathcal{E}(s_k^{*}\oplus g_k^{*}))$
    }
  }

  \tcp{(3) Batch-level post-verification with rollback}
  Rerun the batch $\mathcal{B}_i$ with $\mathcal{M}_i'$ and compute $\mathrm{Acc}_{new}\leftarrow \frac{1}{N}\sum_{k=1}^N R(\pi_{\theta}(z_k|\mathcal M'_i))$ \;
  $\mathcal{M}_{i+1}\leftarrow \mathcal{M}_{i}'$ \textbf{if} $\mathrm{Acc}_{new}\ge \mathrm{Acc}_{base}$ \textbf{else} roll back $\mathcal{M}_{i+1}=\mathcal{M}_{i}$ \;
}
\end{algorithm}

\section{Experiments}
\textbf{Experimental Settings.} We adopt two experimental settings: \ding{182} \textit{Supervised manner.} We hold out the test set for evaluation, and use data sampled from the training split to update and save the external mistake notebook, which is then used during evaluation. \ding{183} \textit{Test-time scaling (TTS).} We directly evaluate on the test set while continuously updating the external memory online, allowing the model to improve progressively as testing proceeds.

\textbf{Evaluation Datasets and Metrics.} We evaluate on six benchmarks covering three vision language domains: STEM/Math (MMMU$_{val}$~\cite{mmmu}, MathVista~\cite{mathvista}), General VQA (MMStar~\cite{mmstar}, RealworldQA~\cite{realworldqa}), and Document OCR (AI2D~\cite{ai2d}, ChartQA~\cite{chartqa}). MMMU and MathVista emphasize multi-discipline reasoning and mathematical problem solving; MMStar and RealworldQA focus on open-world visual understanding; AI2D and ChartQA evaluate diagram/chart comprehension and text-grounded reasoning. Following prior work, we report accuracy (\%) as the primary evaluation metric for all datasets. Among them, MMMU and MathVista are trained in a supervised manner, while the other benchmarks are evaluated using test-time scaling (TTS); please refer to Appendix~\ref{appendix:bench} for details.

\begin{table*}[t]
\centering
\caption{\textbf{Main results.} We report accuracy and average score, together with the estimated training/API cost, the average number of entries in the memory (\textbf{Mem}), and the average number of tokens in the guidance (\textbf{Len}). $\ddagger$ denotes results reproduced by us with a specific answer-parsing strategy, and the best performance is in \textbf{bold.}}
\label{tab:main_results}

\setlength{\tabcolsep}{2pt}
\resizebox{\textwidth}{!}{%
\begin{tabular}{lccccc|cc|ccc}
\toprule
\multirow{2}{*}{\textbf{Method}} &
\multirow{2}{*}{\textbf{Cost}} &
\multirow{2}{*}{\textbf{Mem}} &
\multirow{2}{*}{\textbf{Len}} &
\multicolumn{2}{c|}{\textbf{STEM Puzzle}} &
\multicolumn{2}{c|}{\textbf{General VQA}} &
\multicolumn{2}{c}{\textbf{Document OCR}} & \multirow{2}{*}{\textbf{Avg.}} \\
\cmidrule(lr){5-6}\cmidrule(lr){7-8}\cmidrule(lr){9-10}
& & & &
MMMU$_{val}$ & MathVista &
MMStar & RealworldQA &
AI2D & ChartQA \\

\midrule
\rowcolor{black!5}
\multicolumn{11}{@{}c@{}}{\textit{Self-Evolving: Compare with SFT- and RL-based methods}} \\
Vision-Zero~(\citeauthor{visionzero}) & \multirow{4}{*}{Basically} & -- & -- & 58.8 & 72.6 & 65.2 & 68.5 & 84.5 & 86.3 & 72.65 \\
EvoLMM~(\citeauthor{evolmm}) & \multirow{4}{*}{$>\$500$} & -- & -- & 52.0 & 70.5 & -- & -- & 83.4 & 86.7 & -- \\
iReasoner~(\citeauthor{ireasoner}) &  & -- & -- & 52.4 & 69.7 & -- & -- & 83.9 & 85.8 & -- \\
VisPlay~(\citeauthor{visplay}) &  & -- & -- & 54.9 & 68.2 & 65.1 & 69.0 & -- & 86.2 & -- \\
DPE ~(\citeauthor{dpe}) & & -- & -- & 69.1 & 76.2 & 62.1 & 72.1 & -- & 84.8 & -- \\
\midrule
\rowcolor{black!5}
\multicolumn{11}{@{}c@{}}{\textit{Self-Evolving: Qwen3-VL-8B-Instruct (Open-source)}} \\
Vanilla$^\dagger$ & -- & -- & -- & 67.1 & 73.2 & 62.1 & 73.2 & 80.8 & 89.6 & 74.3 \\
w/ \textbf{M$^2$Note} & $\approx\$3$ & 50 & 261 &
\textbf{68.4}{\color{purple}(\,$\uparrow$1.3\,)} &
\textbf{77.3}{\color{purple}(\,$\uparrow$4.1\,)} &
\textbf{63.9}{\color{purple}(\,$\uparrow$1.8\,)} &
\textbf{75.4}{\color{purple}(\,$\uparrow$2.2\,)} &
\textbf{82.1}{\color{purple}(\,$\uparrow$1.3\,)} & 
\textbf{89.9}{\color{purple}(\,$\uparrow$0.3\,)} &
\textbf{76.2}{\color{purple}(\,$\uparrow$1.9\,)} \\
w/ CoT & -- & -- & -- & 68.8 & 82.9 & 62.9 & 78.0 & 81.3 & 91.7 & 77.6 \\
w/ CoT+\textbf{M$^2$Note} & $\approx\$4$ & 32 & 315 & \textbf{69.4}{\color{purple}(\,$\uparrow$0.6\,)} & \textbf{83.9}{\color{purple}(\,$\uparrow$1.0\,)} & \textbf{64.9}{\color{purple}(\,$\uparrow$2.0\,)} & \textbf{78.8}{\color{purple}(\,$\uparrow$0.8\,)} & \textbf{81.7}{\color{purple}(\,$\uparrow$0.4\,)} & \textbf{91.7} & \textbf{78.4}{\color{purple}(\,$\uparrow$0.8\,)} \\
\addlinespace[2pt]

\midrule
\rowcolor{black!5}
\multicolumn{11}{@{}c@{}}{\textit{Self-Evolving: Qwen3-VL-Plus (Proprietary)}} \\
Vanilla$^\dagger$ & -- & -- & -- & 77.9 & 78.0 & 68.3 & 78.2 & 85.9 & 88.5 & 79.5 \\
w/ \textbf{M$^2$Note} & $\approx\$7$ & 28 & 462 &
\textbf{79.3}{\color{purple}(\,$\uparrow$1.4\,)} &
\textbf{80.0}{\color{purple}(\,$\uparrow$2.0\,)} &
\textbf{69.8}{\color{purple}(\,$\uparrow$1.5\,)} &
\textbf{79.4}{\color{purple}(\,$\uparrow$1.2\,)} &
\textbf{86.7}{\color{purple}(\,$\uparrow$0.8\,)} & 
\textbf{89.2}{\color{purple}(\,$\uparrow$0.7\,)} & 
\textbf{80.7}{\color{purple}(\,$\uparrow$1.2\,)} \\
\addlinespace[2pt]

\midrule
\rowcolor{black!5}
\multicolumn{11}{@{}c@{}}{\textit{Self-Evolving: GPT-5.4 (Proprietary)}} \\
Vanilla$^\dagger$ & -- & -- & -- & 77.1 & 77.2 & 66.5 & 80.8 & 85.6 & 85.6 & 78.8 \\
w/ \textbf{M$^2$Note} & $\approx\$7$ & 28 & 462 &
\textbf{77.9}{\color{purple}(\,$\uparrow$0.8\,)} &
\textbf{78.0}{\color{purple}(\,$\uparrow$0.8\,)} &
\textbf{67.8}{\color{purple}(\,$\uparrow$1.3\,)} &
\textbf{81.8}{\color{purple}(\,$\uparrow$1.0\,)} &
\textbf{86.1}{\color{purple}(\,$\uparrow$0.5\,)} & 
\textbf{85.8}{\color{purple}(\,$\uparrow$0.2\,)} & 
\textbf{79.6}{\color{purple}(\,$\uparrow$0.8\,)} \\
\addlinespace[2pt]

\midrule
\rowcolor{black!5}
\multicolumn{11}{@{}c@{}}{\textit{Cross-Model Evolving: Qwen3-VL-8B-Instruct (Tuning) and Qwen3-VL-Plus (Tuner)}} \\
Vanilla$^\dagger$ & -- & -- & -- & 67.1 & 73.2 & 62.1 & 73.2 & 80.8 & 89.6 & 74.3 \\
w/ \textbf{M$^2$Note} & $\approx\$5$ & 43 & 445 &
\textbf{68.9}{\color{purple}(\,$\uparrow$1.8\,)} &
\textbf{78.3}{\color{purple}(\,$\uparrow$5.1\,)} &
\textbf{64.0}{\color{purple}(\,$\uparrow$1.9\,)} &
\textbf{77.1}{\color{purple}(\,$\uparrow$3.9\,)} &
\textbf{81.0}{\color{purple}(\,$\uparrow$0.2\,)} &
\textbf{90.1}{\color{purple}(\,$\uparrow$0.5\,)} &
\textbf{76.6}{\color{purple}(\,$\uparrow$2.3\,)} \\
\addlinespace[2pt]
\bottomrule
\end{tabular}%
}
\end{table*}

\textbf{Implementation Details.} Our M$^2$Note does not require training model parameters; it only updates a mistake notebook stored as external memory in the JSONL format. We evaluate three VLMs: the open-source \textit{Qwen3-VL-8B-Instruct}~\cite{qwen3-vl}, the closed-source \textit{Qwen3-VL-Plus}~\cite{qwen3-vl} and \textit{GPT-5.4}~\cite{hurst2024gpt}. For multimodal RAG, we use \textit{Qwen3-VL-Embedding}~\cite{qwen3-vl-embedding} as the embedding model. In addition to these models, all ablation variants are also implemented through API requests. Unless otherwise specified, we set the RAG top-$K$ to $K=1$ and the RAG threshold to $\alpha=0.4$. The batch size for both training and inference is $16$. For supervised experiments on MMMU and MathVista, we run $10$ and $20$ training steps, respectively. More details can be found in Appendix~\ref{appendix:imple}.

\subsection{Main Results}
Table~\ref{tab:main_results} reports the main results of M$^2$Note on six multimodal reasoning benchmarks: \ding{182} \textit{Consistent gains across domains and backbones.} M$^2$Note yields consistent improvements on different domains, and the gains hold for both an open-source backbone and a stronger proprietary backbone. \ding{183} \textit{Cost- and sample-efficiency.} Compared to RL-based self-evolving approaches, M$^2$Note achieves competitive or even better performance with dramatically lower cost via pure API requests. Under the same backbone model (\textit{Qwen3-VL-8B-Instruct}), M$^2$Note matches or slightly outperforms DPE~\cite{dpe} while avoiding RL-style training overhead. \ding{184} \textit{Multiple evolving modes.}
M$^2$Note supports both self-evolving (single model acts as both the tuning and tuner models) and cross-model evolving (a stronger tuner model supervises a weaker tuning model), enabling capability transfer without any weight updates. \ding{185} \textit{CoT compatibility.} M$^2$Note can work together with Chain-of-Thought (CoT) prompting~\cite{cot,chain}, leading to further gains over either CoT or M$^2$Note alone. \ding{186} \textit{Compact notebook with interpretable scaling trends.} We observe that stronger backbones make fewer mistakes and require fewer notebook entries (\textbf{Mem}), while a stronger tuner (\textit{Qwen3-VL-Plus}) tends to produce more detailed guidance, resulting in longer length (\textbf{Len}).



\begin{figure*}[t]
    \centering
    \small
    \begin{subfigure}{0.32\textwidth}
        \centering
        \includegraphics[width=\linewidth]{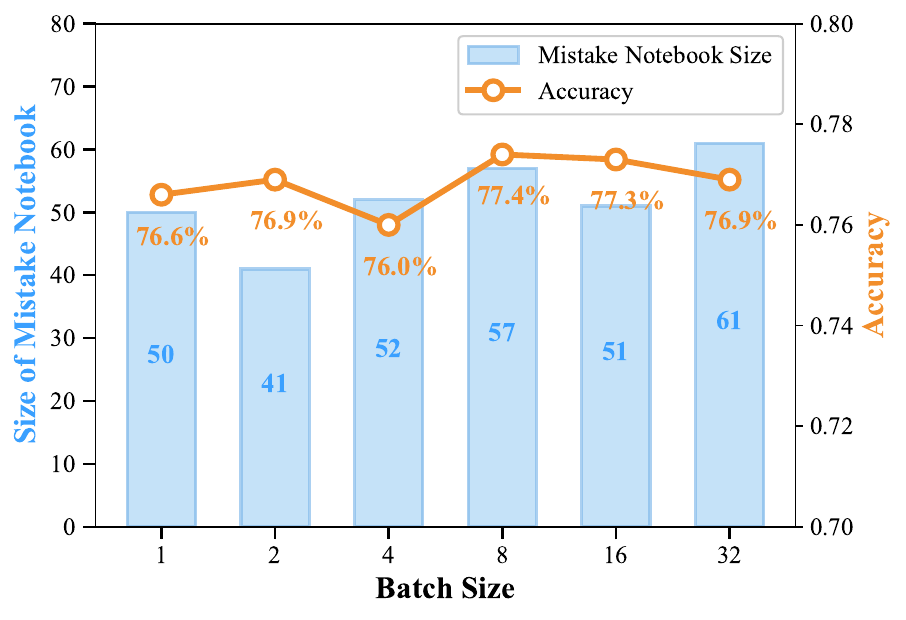}
        \caption{Batch size sensitivity.}
        \label{img:abl_batchsize}
    \end{subfigure}
    \hfill
    \begin{subfigure}{0.31\textwidth}
        \centering
        \includegraphics[width=\linewidth]{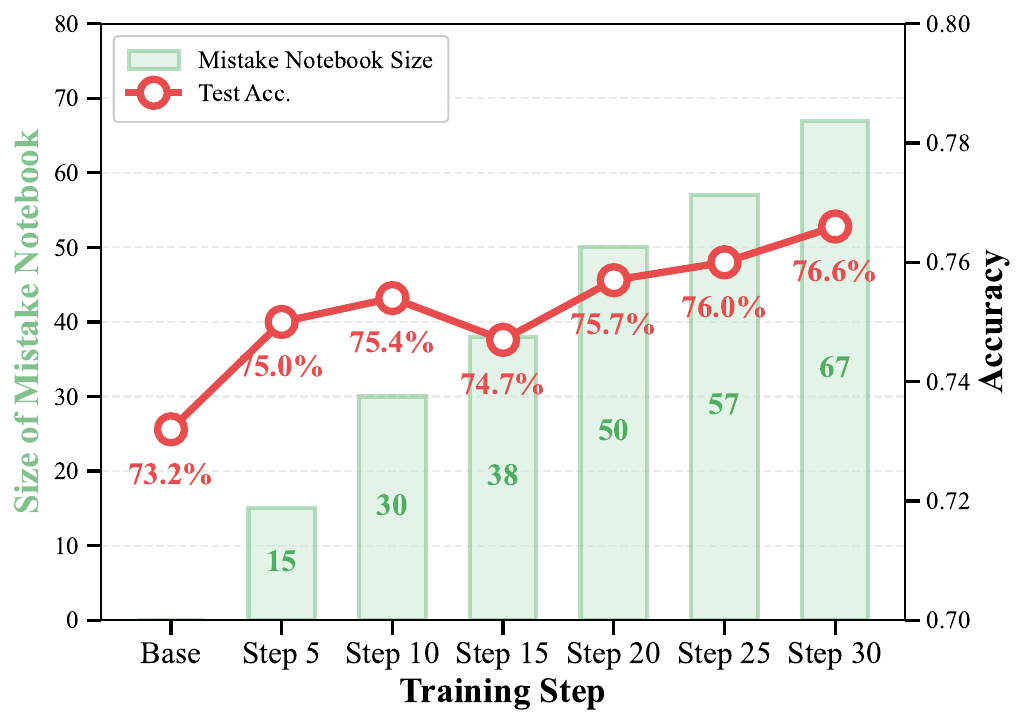}
        \caption{Evolving as training progresses.}
        \label{img:abl_step}
    \end{subfigure}
    \hfill
    \begin{subfigure}{0.35\textwidth}
        \centering
        \includegraphics[width=\linewidth]{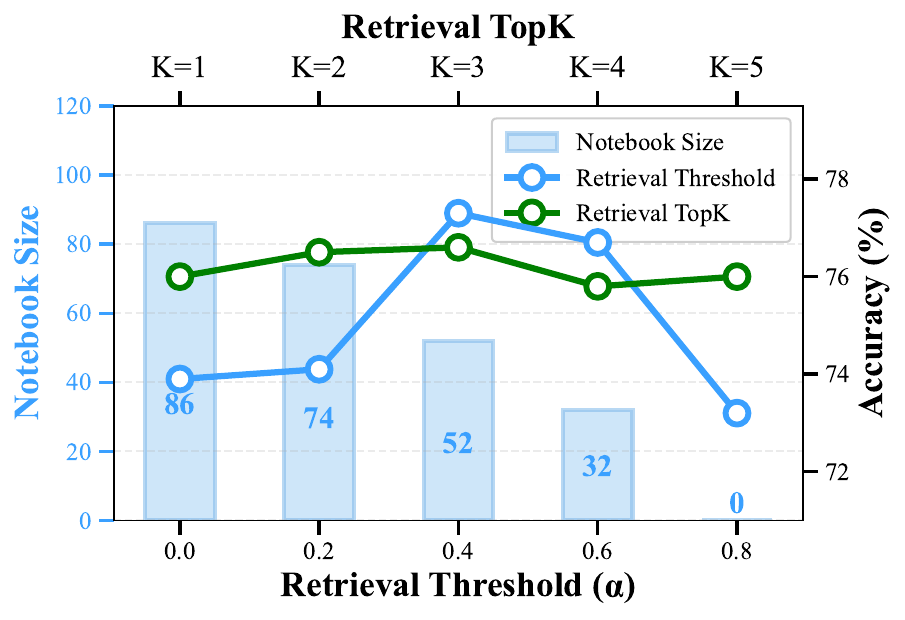}
        \caption{Ablation study of the RAG settings.}
        \label{img:rag}
    \end{subfigure}
    \caption{\textbf{Effects of different hyperparameters on performance and notebook size.} Experiments are conducted on MathVista with \textit{Qwen3-VL-8B-Instruct}.}
    \label{img:hyperparams_rag}
\end{figure*}

\subsection{Ablation Study}
\textbf{Sensitivity Analysis of Hyperparameters.} Results in Figure~\ref{img:abl_batchsize} and \ref{img:abl_step} reveal two trends: \ding{182} M$^2$Note is relatively insensitive to batch size. Across a wide range (1--32), test accuracy remains stable at around 76\%--77.4\%, suggesting that the closed-loop update does not depend heavily on a specific batch size. Moderate batch sizes (e.g., 8 or 16) perform best, likely because they support more robust batch-level memory refinement and make the \textit{verify} stage more reliable. Meanwhile, notebook size does not increase monotonically with batch size (about 41--61 entries), indicating that subject-level merging prevents memory from scaling directly with the batch. \ding{183} As training progresses, the notebook gradually expands (15 to 67 entries), while accuracy improves from 73.2\% to 76.6\%. Although there is some fluctuation around step 15, the overall upward trend suggests that continued evolution accumulates reusable guidance and improves test generalization.

\textbf{Multimodal RAG Settings.} Figure~\ref{img:rag} studies the effect of the retrieval threshold $\alpha$ and the retrieval top-$K$ on MathVista (\textit{Qwen3-VL-8B-Instruct}). Increasing $\alpha$ makes the system more conservative, reducing the number of queries that trigger note insertion; as fewer queries retrieve any guidance, batch performance remains unchanged before and after the update, thus shrinking the final notebook size. In terms of accuracy, a moderate threshold performs best: $\alpha{=}0.4$ achieves the highest accuracy, while overly strict retrieval (e.g., $\alpha{=}0.8$) disables the notebook and degrades performance. We also observe that the system is robust to the retrieval top-$K$, with performance varying only slightly. Overall, we prefer a smaller value ($K{=}1$ by default) to reduce the risk of hallucinations.

\noindent\textbf{Effectiveness of Batch-Level Post-Verification.} Table~\ref{tab:ablation_batch_level} examines the batch-level post-verification, namely \textit{accept-if-improves}. For \textit{Qwen3-VL-8B-Instruct}, enabling post verification yields the best overall results, notably improving MathVista (73.2 $\rightarrow$ 77.3) and MMMU$_{val}$ (67.1 $\rightarrow$ 68.4), showing that filtering noisy updates stabilizes evolution and makes the accumulated notes more reliable. Without post verification, performance becomes less consistent (e.g., a drop on MMMU$_{val}$), indicating that blindly committing proposed notes may introduce harmful or noisy guidance. For the stronger \textit{Qwen3-VL-Plus} backbone, the effect is smaller but still positive on average, suggesting that as the base model makes fewer mistakes, the notebook receives fewer high-impact updates, yet conservative verification remains useful for preventing degradation.

\begin{table}
\centering
\caption{\textbf{Effectiveness of batch-level post-verification.} We report results on three benchmarks across two backbones. Best results are marked in \textbf{bold}.}
\label{tab:ablation_batch_level}
\renewcommand{\arraystretch}{1.1}
\resizebox{\columnwidth}{!}{
\begin{tabular}{lccc}
\toprule
\textbf{Method} & \textbf{MathVista} & \textbf{MMMU$_{val}$} & \textbf{MMStar} \\
\midrule
\rowcolor{black!5}
\multicolumn{4}{@{}c@{}}{\textit{Self-Evolving: Qwen3-VL-8B-Instruct (Open-source)}} \\
\addlinespace[2pt]
Vanilla & 73.2 & 67.1 & 62.1 \\
w/o Batch-Level Post-Verification  & 76.2 & 65.7 & 62.7 \\
\rowcolor{blue!7}
w/ Batch-Level Post-Verification  & \textbf{77.3} & \textbf{68.4} & \textbf{63.9} \\
\midrule
\rowcolor{black!5}
\multicolumn{4}{@{}c@{}}{\textit{Self-Evolving: Qwen3-VL-Plus (Proprietary)}} \\
\addlinespace[2pt]
Vanilla & 78.0 & 77.9 & 68.3 \\
w/o Batch-Level Post-Verification  & \textbf{80.4} & 77.0 & \textbf{69.6} \\
\rowcolor{blue!7}
w/ Batch-Level Post-Verification  & 80.1 & \textbf{79.7} & 69.4 \\
\bottomrule
\end{tabular}
}
\end{table}

\textbf{Ablation on retrieval embeddings.} We study how the embedding model and input modality affect M$^2$Note as shown in Table~\ref{tab:ablation_embedding}. Using a text-only embedding model (\textit{Qwen3-Embedding}) can improve MathVista, but tends to generalize worse on other benchmarks such as MMStar. In contrast, VLM-based embedding models (\textit{Qwen3/2.5-VL-Embedding}) that encode both image and text consistently achieve better performance, suggesting that multimodal indexing better captures the visual cues needed to retrieve relevant notes. This is further supported by the modality ablation: using a single modality, either image or text, is inferior to using both, indicating that visual and textual signals are complementary for robust note retrieval.

\begin{table}
\centering
\caption{\textbf{Impact of different embedding models and retrieval modes.} We report results using \textit{Qwen3-VL-Instruct-8B}. Best results are marked in \textbf{bold}.}
\label{tab:ablation_embedding}
\setlength{\tabcolsep}{2pt}
\renewcommand{\arraystretch}{1.1}
\resizebox{\columnwidth}{!}{
\begin{tabular}{lccccc}
\toprule
\textbf{Embedding Model} & \textbf{Image} & \textbf{Text} & \textbf{MathVista} & \textbf{MMMU$_{val}$} & \textbf{MMStar} \\
\midrule
\rowcolor{black!5}
\multicolumn{6}{@{}c@{}}{\textit{Self-Evolving: Qwen3-VL-8B-Instruct (Open-source)}} \\
\addlinespace[2pt]
Vanilla & - & - & 73.2 & 67.1 & 62.1 \\
\textit{Qwen3-Embedding}~(\citeauthor{qwen3-embedding}) & - & \cmark & 75.5 & 66.6 & 59.1 \\
\textit{Qwen2.5-VL-Embedding}~(\citeauthor{qwen3-vl-embedding}) & \cmark & \cmark & \textbf{77.3} & 66.2 & 63.2 \\
\midrule
\textit{Qwen3-VL-Embedding}~(\citeauthor{qwen3-vl-embedding}) & \cmark & \xmark & 76.3 & 66.0 & 63.0 \\
\textit{Qwen3-VL-Embedding}~(\citeauthor{qwen3-vl-embedding}) & \xmark & \cmark & 77.1 & 66.2 & 62.3 \\
\rowcolor{blue!7}
\textit{Qwen3-VL-Embedding}~(\citeauthor{qwen3-vl-embedding}) & \cmark & \cmark & \textbf{77.3} & \textbf{68.4} & \textbf{63.9} \\
\bottomrule
\end{tabular}
}
\end{table}

\begin{figure*}[t]
\centering
\includegraphics[width=0.9\textwidth]{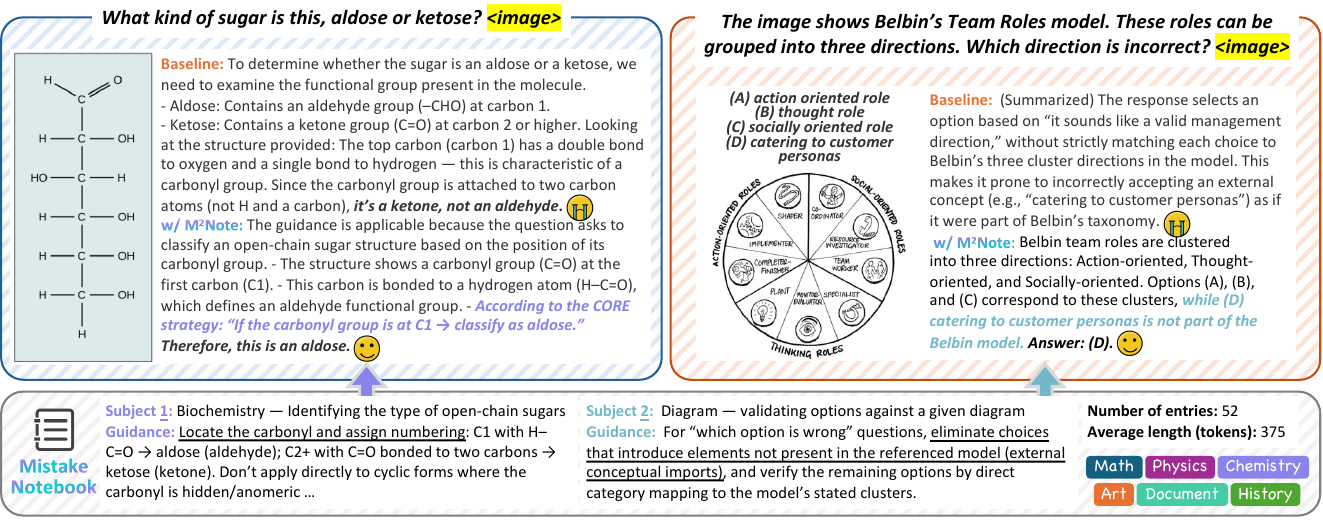}
\caption{\textbf{Qualitative results.} For a STEM (biochemistry) question and a document understanding question, the baseline tends to rely on superficial cues and makes incorrect selections, while M$^2$Note retrieves subject-specific guidance from the external mistake notebook and corrects the reasoning by enforcing key structural checks (e.g., carbonyl position for aldose/ketose; option-to-cluster mapping for Belbin roles).}
\label{imgs:qualitative}
\end{figure*}

\textbf{Notebook transfer across models.} We evaluate whether an evolved mistake notebook can transfer across different VLMs (Table~\ref{tab:generalizable}). Specifically, we first self-evolve a notebook with one backbone (e.g., \textit{Qwen3-VL-8B-Instruct} or \textit{Qwen3-VL-Plus}), then freeze it and use it as external guidance for a different tuning model, without further evolution or parameter updates. Overall, the transferred notebooks remain effective, indicating that M$^2$Note captures partially model-agnostic and reusable guidance. However, cross-model transfer is not always risk-free, and mismatched guidance can hurt performance in some cases. Moreover, compared with self-evolved notebooks in Table~\ref{tab:main_results}, transferred notebooks generally yield smaller gains, suggesting that the strongest improvements come from adapting to each model’s own failure patterns.

\begin{table}
\centering
\caption{\textbf{Notebook transfer across different VLMs.} We report results using three different VLMs.}
\label{tab:generalizable}
\setlength{\tabcolsep}{2pt}
\renewcommand{\arraystretch}{1.1}
\resizebox{\columnwidth}{!}{
\begin{tabular}{lccccc}
\toprule
\textbf{Tuning Model} & \textbf{Mem} & \textbf{Len} & \textbf{MathVista} & \textbf{MMMU$_{val}$} & \textbf{MMStar} \\
\midrule
\rowcolor{black!5}
\multicolumn{6}{@{}c@{}}{\textit{Notebook obtained via self-evolving Qwen3-VL-8B-Instruct}} \\
\addlinespace[2pt]
\textit{Qwen3-VL-Plus}~(\citeauthor{qwen3-vl}) & -- & -- & 78.0 & 77.9 & 68.3 \\
w/ Specific Notebook & 61 & 261 & 78.7{\color{purple}(\,$\uparrow$0.7\,)} & 76.8{\color{blue}(\,$\downarrow$1.1\,)} & 68.6{\color{purple}(\,$\uparrow$0.3\,)} \\
\textit{Qwen3-VL-32B-Instruct}~(\citeauthor{qwen3-vl}) & -- & -- & 81.1 & 73.1 & 67.8 \\
w/ Specific Notebook & 61 & 261 & 81.4{\color{purple}(\,$\uparrow$0.4\,)} & 74.1{\color{purple}(\,$\uparrow$1.0\,)} & 68.1{\color{purple}(\,$\uparrow$0.3\,)} \\
\midrule
\rowcolor{black!5}
\multicolumn{6}{@{}c@{}}{\textit{Notebook obtained via self-evolving Qwen3-VL-Plus}} \\
\textit{Qwen3-VL-8B-Instruct}~(\citeauthor{qwen3-vl}) & -- & -- & 73.2 & 67.1 & 62.1 \\
w/ Specific Notebook & 33 & 469 & 74.7{\color{purple}(\,$\uparrow$1.5\,)} & 67.4{\color{purple}(\,$\uparrow$0.3\,)} & 62.7{\color{purple}(\,$\uparrow$0.6\,)} \\
\addlinespace[2pt]
\bottomrule
\end{tabular}
}
\end{table}

\subsection{Qualitative Results} Fig.~\ref{imgs:qualitative} presents two representative cases illustrating how M$^2$Note improves multimodal reasoning. In the STEM (biochemistry) example on the left, the baseline gives a plausible explanation but misidentifies key structural cues, leading to an incorrect classification. With M$^2$Note, the model retrieves subject-level guidance from the external mistake notebook (e.g., locate the carbonyl and verify its position), which enforces essential structural checks and corrects the conclusion. In the diagram understanding example on the right, the baseline tends to pick an option that sounds reasonable without strictly aligning choices to the diagram’s defined taxonomy, making it vulnerable to introducing hallucinations. M$^2$Note retrieves guidance that explicitly requires option-to-cluster mapping and eliminates elements not present in the reference model, yielding the correct answer. More visualizations, including case analyses and mistake-notebook visualizations, are provided in Appendix~\ref{appendix:vis} (Figure~\ref{img:huanjue}).

\section{Conclusion}
We introduce M$^2$Note, a training-free framework for continually improving VLMs through multimodal mistake notebook learning. Instead of updating model weights, M$^2$Note distills model failures into reusable subject-specific guidance and retrieves relevant notes at inference time via multimodal RAG, helping the model verify its reasoning and avoid repeated errors. A key mechanism is batch-level \textit{accept-if-improves} verification, which filters noisy notebook updates and stabilizes continual evolution by retaining only beneficial changes. Across six benchmarks covering STEM/math reasoning, general VQA, and document understanding, M$^2$Note yields consistent gains in both \textit{self-evolving} and \textit{cross-model evolving} settings, while also combining effectively with CoT prompting. Overall, these results suggest that reusing verified failure cases provides a practical and effective way to enhance VLM robustness at deployment.

\section{Limitation}
M$^2$Note evolves in a way that resembles how humans learn from experience: instead of rewriting one’s “brain” (model weights), it repeatedly summarizes failures into reusable rules and consults them when facing similar situations. This mechanism is most effective when mistakes exhibit recurring structure, so that a distilled note can be reliably reused. As a result, M$^2$Note tends to work best in relatively narrow domains (e.g., math reasoning), where tasks share stable abstractions and retrieved guidance is more likely to transfer. In broad domains with long-tail visual diversity, relevant notes can be harder to retrieve, and mismatched guidance may introduce misleading context or amplify hallucinations, making continual evolution less stable. These limitations suggest two directions: \ding{182} building more generalizable notebooks through stronger abstraction and a more reliable post-update verification strategy, so that notes transfer beyond near-duplicate failures and harmful guidance is less likely to be retrieved; and \ding{183} enriching notes beyond text (e.g., reference images, tool-use traces, structured checklists) to provide more grounded and actionable guidance.

\bibliography{custom}

\newpage
\appendix
\setcounter{section}{0}
\renewcommand{\thesection}{\Alph{section}}
\section{Prompts Used in the M$^2$Note Framework}
\label{appendix:prompt}
\begin{figure*}[t]
\centering
\includegraphics[width=1\textwidth]{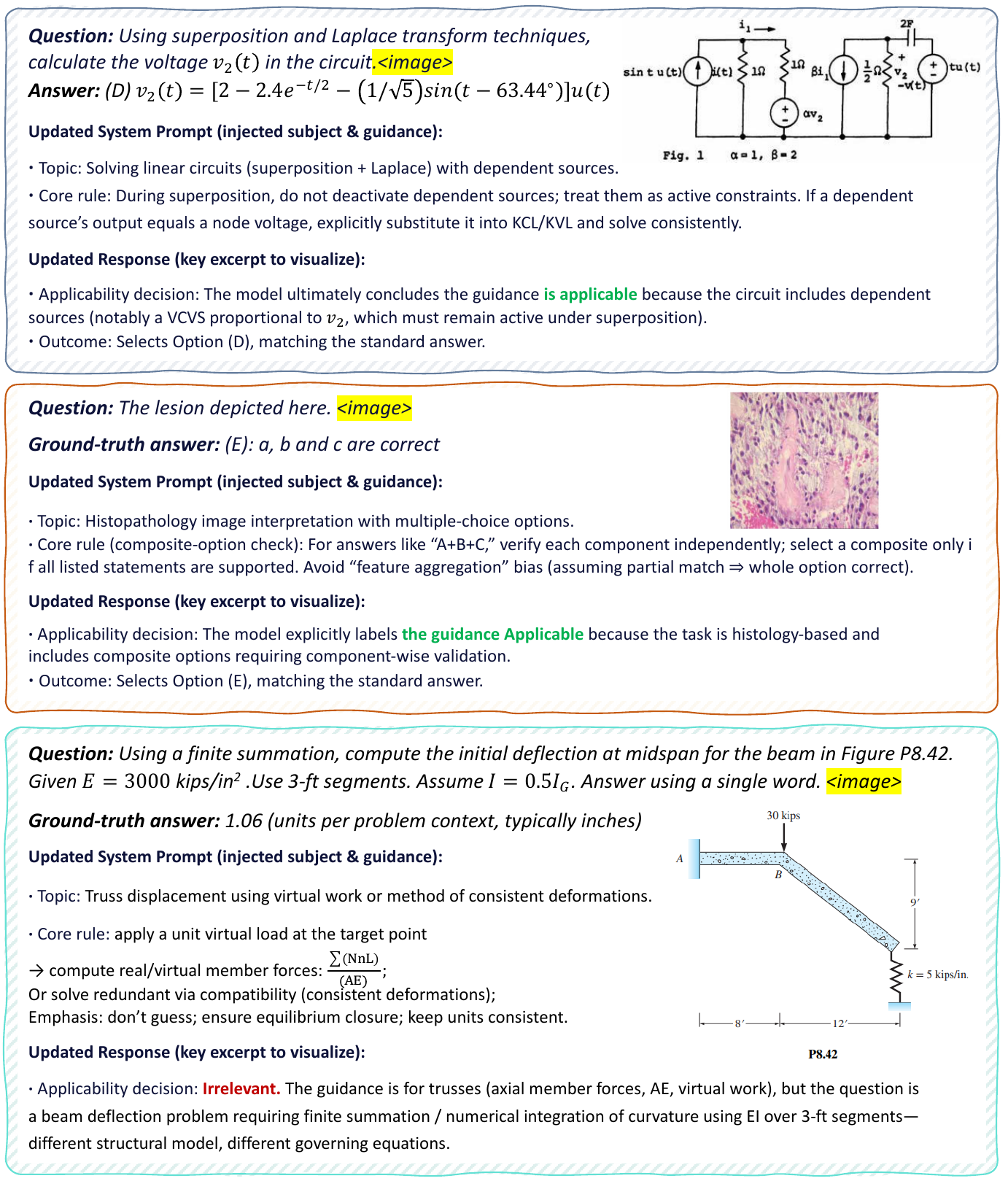}    
\caption{\textbf{Qualitative examples of applicability judgment for retrieved guidance}. The model explicitly decides whether the injected subject-guidance pair is applicable: the first two cases are \textit{Applicable} and lead to correct answers, while the last case is \textit{Irrelevant} and is rejected.}
\label{img:huanjue}
\end{figure*}

\subsection{Tuner Model Prompts}
\begin{promptbox}{Guidance Extraction}
You are a Cognitive Strategy Expert. Your goal is to induce \textbf{generalizable rules} from specific model errors, not just correct them.

$~$

Subject: \{subject\}

Error Data: \{error\_context\}

$~$

Task: Analyze the underlying logical flaws in the Error Data and formulate a reusable problem-solving heuristic.

$~$

Constraints (Total $\leq$ 350 words):

1. \textbf{Abstracted Error Pattern} ($\leq$ 60 words)
   
   - Describe the \textit{type} of situation where this error occurs (strip specific numbers/names).
   
   - Do NOT simply rewrite the original questions.

2. \textbf{Root Cause Analysis} ($\leq$ 50 words)
   
   - Identify the cognitive gap (e.g., confusing correlation with causation, ignoring boundary conditions).

3. \textbf{Generalizable Strategy (CORE)} ($\leq$ 120 words)
   
   - Provide a step-by-step heuristic or checklist applicable to ANY similar future problem.
   
   - Use “If [condition], Then [action]” format where possible.
   
   - Must work beyond the provided examples.

4. \textbf{Anti-Patterns \& Boundary} ($\leq$ 80 words)
   
   - List 1-2 specific scenarios where this strategy should NOT be applied.
  
   - List 1-2 common misinterpretations of this guidance.

$~$

Output Style:

- High-level, structured, knowledge-focused, and reusable for similar future questions.

- Avoid referencing specific details from the provided Error Data unless illustrating a pattern.

\end{promptbox}

\newpage
\begin{promptbox}{Subject Classification}
You are an expert in categorizing questions into precise, high-relevance subjects for Retrieval-Augmented Generation (RAG).

$~$

Your goal is to assign each question a subject label (about 10-50 words) that:

- Maximizes retrieval relevance by precisely describing the problem type and solution method.

- Groups only genuinely similar questions together (same domain AND same approach).

- Avoids over-broad categories that would match unrelated problems.

- Reuses the same subject name for closely related questions.

$~$

CRITICAL: The subject must be specific enough to prevent irrelevant retrieval. Include, when applicable:

1. Primary Domain (e.g., Combinatorics, Complex Analysis, Linear Algebra, Physics, Programming, Document Understanding)

2. Problem Type (e.g., counting with constraints, roots of unity products, debugging API parameters, OCR table extraction)

3. Solution Method (e.g., stars and bars, polynomial/root identities, Hensel's lemma, reproduce-minimize-fix)

$~$

Examples of GOOD subjects (specific):

\cmark Complex Analysis: Evaluating products over roots of unity using polynomial evaluation and complex identities

\cmark Document Understanding: Extracting tables from scanned PDFs using layout detection + OCR + row/column reconstruction

$~$

Examples of BAD subjects (too broad):

\xmark number theory (too broad - could match any modulo problem)

\xmark physics (too broad - could match any physics problem)

$~$

Key principle: If a problem spans multiple domains, label it by the primary one.

$~$

Output only the finalized subject label(s).
\end{promptbox}

\begin{promptbox}{Guidence Merge}
Synthesize guidance for subject: \{subject\}

Existing guidance from related subjects: \{existing\_guidance\}

New guidance: \{new\_guidance\}

$~$

Merge into a single coherent guidance (max 2048 chars) that:

- Combines insights from related subjects with new guidance

- Eliminates redundancy while preserving key information and examples of the mistakes

- \textbf{Preserves and emphasizes applicability conditions} 

- clearly state when each method applies

- Focuses on actionable advice

- Maintains consistent style

- \textbf{Includes warnings about when NOT to apply the guidance} to avoid misapplication

$~$

Merged guidance:
\end{promptbox}

\begin{promptbox}{Subject Merge}
The following subjects are related:
\{subjects\}

Provide a single, concise subject name ($\leq$10 words) that best represents all of these. Only output the subject name, nothing else.
\end{promptbox}

\subsection{Tuning Model Prompts}
\label{appendix:hallu}
\begin{promptbox}{User Prompt}
Question: \{question\}

Options: \{options\}

Please select the correct answer from the options above.
\end{promptbox}

\newpage
\begin{promptbox}{RAG-Enhanced System Prompt}
The following mistake notes are not necessarily tied to the current question, but you may use them to deepen your analytical approach

$~$

\textbf{IMPORTANT}: Before applying any guidance below, carefully evaluate:

1. Does the current problem match the applicability conditions stated in the guidance?

2. Is the problem type and context similar to the examples in the guidance?

3. If the problem is totally different (e.g., combinatorics vs modulo arithmetic, complex numbers vs number theory), do NOT force-fit the guidance.

4. Only use guidance that is clearly relevant to the current problem.

$~$

\textbf{Gudience Subject}: \{subject\} - \{guidance\}

Before solving, review the guidance. State whether it is: applicable, partially applicable, or irrelevant. Use only applicable parts.

\end{promptbox}

Since the guidance introduced via multimodal RAG~\cite{rag,mrag} may not necessarily apply to the current question, in addition to the top-$K$ and threshold settings in the RAG configuration, we further emphasize in the system prompt that the model should judge whether the retrieved content is applicable to the current question. This is crucial for reducing hallucinations. As shown in Fig.~\ref{img:huanjue}, several cases are presented where the model’s response explicitly includes this judgment: the first two cases are applicable, while the last one illustrates an inapplicable scenario. Table~\ref{tab:huanjue} compares blindly following the guidance (naive) with introducing a judgment mechanism (judgement); the latter effectively improves the model’s problem-solving accuracy.

\begin{table}[t]
\centering
\caption{\textbf{Effect of applicability judgment for retrieved guidance}. Results on three benchmarks and two backbones. \textit{Naive} uses retrieved guidance directly, while \textit{Judgment} verifies relevance first, consistently improving performance. Best results are in \textbf{bold}.}
\label{tab:huanjue}
\renewcommand{\arraystretch}{1.1}
\resizebox{\columnwidth}{!}{
\begin{tabular}{lccc}
\toprule
\textbf{Method} & \textbf{MathVista} & \textbf{MMMU$_{val}$} & \textbf{MMStar} \\
\midrule
\rowcolor{black!5}
\multicolumn{4}{@{}c@{}}{\textit{Self-Evolving: Qwen3-VL-8B-Instruct (Open-source)}} \\
\addlinespace[2pt]
Vanilla & 73.2 & 67.1 & 62.1 \\
w/ M$^2$Note (Naive) & 76.2 & 67.8 & \textbf{64.1} \\
\rowcolor{blue!7}
w/ M$^2$Note (Judgment)  & \textbf{77.3} & \textbf{68.4} & 63.9 \\
\midrule
\rowcolor{black!5}
\multicolumn{4}{@{}c@{}}{\textit{Self-Evolving: Qwen3-VL-Plus (Proprietary)}} \\
\addlinespace[2pt]
Vanilla & 78.0 & 77.9 & 68.3 \\
w/ M$^2$Note (Naive) & 79.5 & 76.9 & 69.0 \\
\rowcolor{blue!7}
w/ M$^2$Note (Judgment) & \textbf{80.1} & \textbf{79.7} & \textbf{69.4} \\
\bottomrule
\end{tabular}
}
\end{table}

\section{More Visualizations}
\label{appendix:vis}

Below we present several entries from the mistake notebook, obtained by training \textit{Qwen3-VL-8B-Instruct} as the \textbf{Tuning Model} on MMMU~\cite{mmmu}. As shown, the structured guidance is produced using the \textbf{Tuner Model} with the “Guidance Extraction” prompt, while the \textbf{Note} and the \textbf{Task} field are derived using the “Subject Classification” prompt.

\begin{casebox}{teal}{Note 1: Document Understanding}
\field{Task} Extract structured brainstorming workflow steps from a quadrant diagram (visual category mapping + label extraction).

\field{Abstracted Error Pattern}
Options that sound reasonable are mistaken as valid steps even when they contradict the diagram’s collaborative workflow (e.g., silent reading, during disagreement).

\field{Root Cause}
Plausibility is prioritized over fidelity to the visual framework; behavioral/structural constraints are ignored.

\field{Generalizable Strategy (CORE)}
\begin{itemize}
  \item Reject options introducing elements absent from the visual model (silence, conflict resolution, external tools).
  \item Reject behaviors inconsistent with the interaction pattern (solo reading vs group discussion).
  \item If ambiguous, validate against core stages: categorization $\rightarrow$ discussion $\rightarrow$ selection.
  \item Prefer behavioral alignment over fluent wording.
\end{itemize}

\field{Boundary}
Not suitable when silent reflection is explicitly part of the process, or when the model allows multiple valid interpretations.
\end{casebox}

\newpage
\begin{casebox}{purple}{Note 2: Medical Imaging}
\field{Task} Identify anatomical landmarks in fundus photographs (visual cue detection + retinal structure recognition).

\field{Abstracted Error Pattern}
Structural sparsity is overgeneralized into complete absence across a region without checking the exact boundary/layer.

\field{Root Cause}
Confusing minimal presence with absence; weak boundary checking for layered anatomy and region-specific exceptions.

\field{Generalizable Strategy (CORE)}
\begin{itemize}
  \item Verify absence at the exact indicated location (not nearby areas).
  \item Check layers independently: photoreceptors $\rightarrow$ bipolar $\rightarrow$ ganglion $\rightarrow$ blood vessels.
  \item Choose “all of the above” only if every listed condition holds.
  \item Cross-reference standard anatomy for region-specific exceptions.
\end{itemize}

\field{Boundary}
Not for dynamic processes (disease progression) or non-structural properties (function).
\end{casebox}

\begin{casebox}{orange}{Note 3: Chemistry}
\field{Task} Determine the maximum number of saccharin molecules from an atomic inventory (limiting-reactant style stoichiometry).

\field{Abstracted Error Pattern}
Users misidentify the limiting element by ignoring stoichiometric ratios or miscounting atoms in the molecular formula.

\field{Root Cause}
Atom availability is mistaken for molecule yield; constraints are not computed per element to find the true bottleneck.

\field{Generalizable Strategy (CORE)}
\begin{enumerate}
  \item Derive the exact molecular formula (count all atoms).
  \item For each element: available atoms $\div$ required per molecule $\Rightarrow$ max molecules.
  \item The smallest quotient determines the maximum yield (the limiting element).
  \item Sanity-check: if quotients look inconsistent, recheck formula/inventory.
  \item Confirm the final answer satisfies the limiting element’s requirement exactly.
\end{enumerate}

\field{Boundary}
Not for non-stoichiometric settings (catalysis, equilibrium-limited yield) or inventories with impurities/intermediates not modeled.
\end{casebox}

\begin{casebox}{cyan}{Note 4: Education}
\field{Task} Identify a classroom activity for literary analysis via role-play and multi-perspective retelling (media simulation).

\field{Abstracted Error Pattern}
The activity is misidentified by matching the image’s media appearance (e.g., a news set) rather than its pedagogical function.

\field{Root Cause}
Surface-level visual similarity overrides functional interpretation (real media format vs simulated classroom role-play).

\field{Generalizable Strategy (CORE)}
\begin{itemize}
  \item Eliminate options that are literal media formats (e.g., News Program).
  \item Prefer terms describing pedagogical function: interviewing, role assumption, perspective simulation.
  \item Check definitions for student agency, dialogue, and interpretive framing.
  \item Treat broadcast-studio visuals as distractors; focus on the learning objective.
  \item Differentiate close terms: Hot Seat (structured questioning) vs Readers Theatre (scripted reading).
\end{itemize}

\field{Boundary}
Not for purely observational tasks (watching media) or when the explicit goal is media production.
\end{casebox}

\begin{casebox}{green}{Note 5: Document Understanding}
\field{Task} Count pathogens on leaf images (visual segmentation + pattern recognition).

\field{Abstracted Error Pattern}
Multiple symptom patterns are incorrectly treated as multiple pathogens, ignoring overlap and single-agent variability.

\field{Root Cause}
Symptoms (markers) are conflated with causal entities; overlap/co-infection and environmental confounds are not checked.

\field{Generalizable Strategy (CORE)}
\begin{itemize}
  \item Check whether different patterns overlap spatially/temporally; overlap may indicate one agent.
  \item Compare with known symptom profiles to validate distinct causes.
  \item If distinctness is supported, count independent infection zones (not symptom types).
  \item Consider context (treatment, stressors) that can mimic/mask infections.
  \item If evidence is weak, prefer “unknown” over forcing a precise count.
\end{itemize}

\field{Boundary}
Not when symptoms clearly indicate unrelated agents (e.g., fungal + viral), or when the question asks for types rather than counts.
\end{casebox}

\begin{casebox}{red}{Note 6: Combinatorics}
\field{Task} Schedule exams under conflict constraints using graph coloring (minimum number of time slots).

\field{Abstracted Error Pattern}
Minimum slots are underestimated by ignoring constraint propagation or assuming a locally good coloring is globally feasible.

\field{Root Cause}
Chromatic number is conflated with clique size; indirect (non-local) conflicts are not fully enforced/validated.

\field{Generalizable Strategy (CORE)}
\begin{enumerate}
  \item Build a conflict graph: exams = vertices; edges = shared students/resources.
  \item Color the graph (greedy or exact), enforcing all pairwise conflicts.
  \item Validate the schedule: no student may have two exams in the same slot/day.
  \item If violations occur, re-solve using backtracking/constraint propagation.
  \item Re-check feasibility against real-world rules (concurrency constraints).
\end{enumerate}

\field{Boundary}
Not for dynamic/time-dependent constraints beyond a static conflict graph, or when intra-day sequencing/rooms must be modeled.
\end{casebox}

\section{Experimental Settings}
\subsection{Implementation Details}
\label{appendix:imple}
Our supervised training settings on MMMU~\cite{mmmu} and MathVista~\cite{mathvista} are as follows: \ding{182} On MMMU, we train on the official \textit{dev} split (150 STEM questions) to obtain a tuned memory, which is then used as an external notebook during inference on the MMMU \textit{validation} split, where we report accuracy. \ding{183} On MathVista, we randomly sample 320 examples from the \textit{test} split to train the notebook, and evaluate on the \textit{text-mini} split, following common practice in prior work.

For the other benchmarks (MMStar~\cite{mmstar}, AI2D~\cite{ai2d}, RealworldQA~\cite{realworldqa}, and ChartQA~\cite{chartqa}), we adopt a test-time scaling (TTS)~\cite{tts} setup, where the memory is updated online during evaluation to progressively enhance the model’s capability.

\subsection{Benchmark Details}
\label{appendix:bench}
For comprehensive evaluation across diverse visual modalities, reasoning skills, and knowledge domains, we conduct experiments on six widely-used multimodal benchmarks: MMMU~\cite{mmmu}, MathVista~\cite{mathvista}, MMStar~\cite{mmstar}, RealworldQA~\cite{realworldqa}, AI2D~\cite{ai2d}, and ChartQA~\cite{chartqa}. Together, these benchmarks cover broad subject understanding, visual mathematical reasoning, real-world recognition and commonsense, diagram interpretation, and chart comprehension.

\textbf{MMMU}~\cite{mmmu} is a large-scale benchmark designed to measure multimodal understanding across a wide range of academic disciplines and professional fields. Questions typically require jointly interpreting images (e.g., figures, tables, diagrams, screenshots) and text, and then performing domain-specific reasoning. These questions span 30 subjects and 183 subfields, comprising 30 highly heterogeneous image types, such as charts, diagrams, maps, tables, music sheets, and chemical structures. MMMU is challenging due to its breadth of subjects and its emphasis on knowledge-intensive and reasoning-intensive problems rather than purely perceptual recognition.

\textbf{MathVista}~\cite{mathvista} focuses on visual mathematical reasoning, where models must solve math problems grounded in visual context. Inputs often contain charts, diagrams, geometry figures, or real-world scenes with quantitative cues. The benchmark evaluates capabilities such as extracting numerical information from images, mapping visual elements to formal quantities, performing multi-step reasoning, and producing the final mathematical answer. It is well-suited for analyzing error patterns related to perception-to-symbol grounding and step-by-step quantitative reasoning.

\textbf{MMStar}~\cite{mmstar} is a general multimodal evaluation benchmark aimed at assessing robust vision-language understanding across a variety of everyday and document-style visual inputs. Questions span recognition, attribute reasoning, spatial reasoning, and higher-level comprehension. Compared with subject-focused benchmarks, MMStar is commonly used to test overall multimodal competence and generalization, making it useful for verifying whether our method improves broad visual-text reasoning rather than overfitting to a narrow domain.

\textbf{RealworldQA}~\cite{realworldqa} targets question answering in real-world visual scenarios, emphasizing practical knowledge and reasoning grounded in natural images. Questions often rely on recognizing objects, scenes, affordances, and context, and may require commonsense or everyday knowledge beyond simple identification. This benchmark is helpful for evaluating whether the model’s “mistake-driven” updates improve robustness in realistic settings where visual ambiguity and implicit assumptions are common.

\textbf{AI2D}~\cite{ai2d} is a benchmark for science diagram understanding. It contains elementary and middle-school style diagrams (e.g., life cycles, physics illustrations, anatomy/biology schematics) paired with questions that require interpreting labeled components, arrows, processes, and relationships. AI2D is particularly challenging because correct answers often depend on understanding diagram structure and semantics, not just recognizing visual entities.

\textbf{ChartQA}~\cite{chartqa} evaluates chart and plot understanding, including bar charts, line charts, pie charts, and other visualizations commonly seen in reports. Questions require reading values from axes/legends, comparing trends, performing simple arithmetic, or synthesizing information across multiple chart elements. ChartQA is valuable for analyzing errors related to visual-text alignment (e.g., legend-to-series mapping), numerical extraction, and compositional reasoning over structured visual data.

\end{document}